\documentclass[pra,superscriptaddress,twocolumn]{revtex4}
\usepackage{enumerate}
\usepackage{amsfonts,amssymb,amsmath}
\usepackage[]{graphics,graphicx,epsfig}
\usepackage{amsthm}

\usepackage{marvosym}

\usepackage{fdsymbol}
\usepackage{graphicx}
\usepackage{dcolumn}
\usepackage{natbib}
\usepackage{color}
\usepackage{multirow}
\usepackage{ulem}
\newcommand{\ket}[1]{|{#1}\rangle}
\newcommand{\bra}[1]{\langle{#1}|}

\begin{document}


\title{Relative homotopy approach to topological phases in quantum walks}

\author{Andrzej Grudka}
\affiliation{Institute of Spintronics and Quantum Information, Faculty of Physics, Adam Mickiewicz University, 61-614 Pozna\'n, Poland}

\author{Marcin Karczewski}
\affiliation{International Centre for Theory of Quantum Technologies, University of Gdańsk,
80-309 Gda{\'n}sk, Poland}

\author{Pawe{\l} Kurzy{\'n}ski}
\affiliation{Institute of Spintronics and Quantum Information, Faculty of Physics, Adam Mickiewicz University, 61-614 Pozna\'n, Poland}

\author{Jan W{\'o}jcik}
\affiliation{Faculty of Physics, Adam Mickiewicz University, 61-614 Pozna\'n, Poland}

\author{Antoni W{\'o}jcik}
\affiliation{Institute of Spintronics and Quantum Information, Faculty of Physics, Adam Mickiewicz University, 61-614 Pozna\'n, Poland}

\date{\today}


\begin{abstract}
Discrete-time quantum walks (DTQWs) provide a convenient platform for a realisation of many topological phases in noninteracting systems. They often offer more possibilities than systems with a static Hamiltonian. Nevertheless, researchers are still looking for DTQW symmetries protecting topological phases and for definitions of appropriate topological invariants. Although majority of DTQW studies on this topic focus on the so called split-step quantum walk, two distinct topological phases can be observed in more basic models.  Here we infer topological properties of the basic DTQWs directly from the mapping of the Brillouin zone to the Bloch Hamiltonian. We show that for translation symmetric systems they can be characterized by a homotopy relative to special points. We also propose a new topological invariant corresponding to this concept. This invariant indicates the number of edge states at the interface between two distinct phases.
\end{abstract}

\maketitle


\textit{Introduction}.---The importance of topological concepts for understanding electronic  properties in solid state was found about forty years ago. This discovery gave rise to a new field of research  \cite{Kosterlitz:2017vt,Haldane:2017tt,Hasan:2010vc}. It became natural to address topology in various translation invariant systems possessing Bloch band structure. In particular, topological properties were investigated for ultra-cold atoms in optical lattices \cite{Cooper:2019tr,Goldman:2016tw,Zhang:2018uz}, photonics \cite{Ozawa:2019us,Khanikaev:2013vy,Khanikaev:2017ut,Lu:2014us,Smirnova:2020ty}, and even for mechanical and accoustic systems \cite{Ma:2019tr}.

A decade ago it was realized that periodically driven systems (Floquet systems), in particular discrete-time quantum walks (DTQWs) \cite{Kempe:2003aa}, provide a useful platform for studying topological phenomena \cite{Harper:2020to,Kitagawa:2012wa,Kitagawa:2010vm}. Topological aspects of DTQWs have been studied both, theoretically and experimentally (see \cite{Wu:2019vs} for a review). The corresponding experiments can be implemented in various physical setups, e.g., ion traps \cite{Schmitz:2009wr,Zahringer:2010te}, superconducting systems \cite{Flurin:2017vp,Yan:2019ve}, NMR \cite{Du:2003vg,Ryan:2005wp}, optical lattices \cite{Karski:2009ty,Dadras:2018vo}, and linear optics \cite{Do:2005wa,Cardano:wm,Perets:2008tt,Tang:ub}. Topological properties of DTQWs have been studied in 1D \cite{Kitagawa:2010vm,Lam:2016wa,Asboth:2014up,Obuse:2011vf,Asboth:2012ub,Obuse:2015uv,Asboth:2013aa,Chen:2017tq,Moulieras:2013vc,Peng:2021wk,Rakovszky:2015um,Tarasinski:2014tv,Barkhofen:2017uu,Cardano:2016aa,Cardano:2017aa,Cedzich:2018vu,Cedzich:2021tf,Cedzich:2022to,Cedzich2018completehomotopy,Wang:2019tx,Ramasesh:2017uo}, 2D \cite{Kitagawa:2010vm,PhysRevB.99.214303,Asboth:2015we,Edge:2015ww,Mochizuki:2020wr} and even 3D \cite{Panahiyan:2021ui,Khazali2022discretetimequantum} scenarios. In particular, the so called split-step model was investigated theoretically \cite{Asboth:2012ub,Asboth:2013aa,Asboth:2015we,Chen:2017tq,Edge:2015ww,Kitagawa:2010vm,Mochizuki:2020wr,Moulieras:2013vc,Obuse:2015uv,Panahiyan:2021ui,Peng:2021wk,Rakovszky:2015um,Tarasinski:2014tv} and experimentally \cite{Flurin:2017vp,Kitagawa:2012ur,Nitsche:2019vc,Xie:2020wu,Xu:2020wr}. Surprisingly, the simplest DTQWs, to which we refer in this work as the basic DTQWs, do manifest topological properties too \cite{Asboth:2014up,Kitagawa:2010vm,Asboth:2012ub,Obuse:2011vf,Obuse:2015uv,Lam:2016wa}. However, the origin of these topological properties is still elusive and it is particularly important to find symmetries that protect them. 

In this Letter we aim to resolve this issue. We show that  topological phases can be properly identified using the notion of relative homotopy. We use this notion to propose a new topological invariant that indicates the number of edge states at the interface between the phases. We prove that these edge states are protected by the particle-hole symmetry for a large class of basic DTQWs. Finally, we present the exact form of edge states occurring at a sharp interface between distinct topological phases.


\textit{Model}.---The basic DTQW models a dynamics of a one-dimensional particle, a walker, in a discrete space. The state of the walker is described by two degrees of freedom, position $x\in{\mathbf Z}$ and a two-level system -- the coin $c\in \{\rightarrow,\leftarrow\}$. The corresponding Hilbert space has a tensor product structure $\mathcal{H}_d \otimes \mathcal{H}_2$. At any given time the state of the walker is given by 
\begin{equation}\nonumber
    \ket{\Psi(t)} = \sum_x \ket{x}\otimes(a_x(t)\ket{\rightarrow}+b_x(t) \ket{\leftarrow}).
\end{equation}
The evolution of the system is discrete and its one step is generated by a unitary operator $U$
\begin{equation}\nonumber
    \ket{\Psi(t+1)} = U \ket{\Psi(t)},
\end{equation}
whose form is $U = SC$.
In the above 
\begin{equation}\nonumber
S = \sum_x \left(|x+1\rangle\langle x| \otimes \begin{pmatrix}1&0\\0&0\end{pmatrix} + |x-1\rangle\langle x| \otimes \begin{pmatrix}0&0\\0&1\end{pmatrix} \right),
\end{equation}
is the conditional translation, whereas 
\begin{equation}\nonumber
    C = \sum_x \ket{x}\bra{x}\otimes \bar{C}_x,
\end{equation}
is the coin operator. Here we use a bar to denote that a respective operator acts in the coin Hilbert space $\mathcal{H}_2$. $\bar{C}_x$ is an arbitrary $2\times 2$ unitary operator, however we restrict its dependence on $x$ to the following form
\begin{equation}\nonumber
    \bar{C}_x = e^{-i\delta} \begin{pmatrix}\cos\theta_x  \ e^{i\alpha} & \sin\theta_x  \ e^{i(\alpha+\beta)} \\ -\sin\theta_x  \ e^{-i(\alpha+\beta)} & \cos\theta_x  \ e^{-i\alpha}\end{pmatrix}\in \mathcal{U}(2).
\end{equation}
The effective Hamiltonian $H$ is defined as $U = e^{-i H}$.
The eigenvalues of $H$, to which we refer as qusienergies $\omega$, are defined only up to an additive constant $2\pi$. We restrict them to the so called first Floquet zone $-\pi \leq \omega \leq \pi$. In the following sections we recall the DTQW symmetries. 

\textit{Sublattice symmetry (SUB)}.---Single step dynamics is restricted to nearest neighbours only, which implies a natural bipartite structure of the lattice. Note that this SUB  is associated with the evolution operator $U$, and not with the corresponding Hamiltonian $H$.
It follows that the unitary operator
\begin{equation}\nonumber
    \Lambda = \sum_x (-1)^{x}\ket{x}\bra{x} \otimes \bar{I}
\end{equation}
fulfills $\Lambda U\Lambda^{-1}= -U$.
Therefore, if  $\ket{\Psi}$ is an eigenstate of $H$, then $\Lambda\ket{\Psi}$ is also an eigenstate of $H$ and the corresponding quasienergies differ by $\pi$. 

\textit{Particle-hole symmetry (PHS)}.--- PHS is the usual name for an antiunitary operator $\Omega$
fulfilling $\Omega U\Omega^{-1}=U$. Majority of research on topological properties of one-dimensional DTQWs focuses on real coin operators with $(\delta,\alpha,\beta)=(0,0,0)$, or $(\delta,\alpha,\beta)=(\frac{\pi}{2},\frac{\pi}{2},0)$, for which PHS yields $\Omega=K$, where $K$ is the complex conjugation in the position and $\sigma_z$ basis. Note that $\delta \neq 0$ causes a quasienergy shift, in which case one can use a generalized PHS condition $\Omega U\Omega^{-1}  = U'$, where $U'$ is equivalent to $U$ up to a phase factor.

Let us show that basic DTQWs do not require real coin operators to exhibit PHS. Let $U^{(0)}$ be a DTQW operator corresponding to $(\alpha,\beta)=(0,0)$ and let $U$ be a general DTQW operator. We redefine the amplitudes, such that 
\begin{equation}\nonumber
    \ket{\Psi(t)} = \sum_x \ket{x}\otimes(a'_x(t) e^{i \alpha x}\ket{\rightarrow}+b'_x(t) e^{i \alpha x}e^{-i \beta}\ket{\leftarrow}).
\end{equation}
Next, we define a unitary operator
\begin{equation}\nonumber
    W = \left( \sum_x e^{i\alpha x } \ket{x} \bra{x} \right)\otimes \begin{pmatrix}1&0\\0&e^{-i\beta}\end{pmatrix}
\end{equation}
to obtain $ W U^{(0)} W^{-1} = U$.  Therefore, the evolution operator $U$, with a general complex coin, is unitarily equivalent to $U^{(0)}$.

It follows that an antiunitary operator $\Omega = W^2 K$ ($\Omega^2=I$) is the PHS of the DTQW generated by $U$. Moreover, $U = \Omega U \Omega^{-1}$ and $\Omega H\Omega^{-1}=-H$. Thus,
the eigenstates of $H$ form dublets $\{ \ket{\Psi},\Omega \ket{\Psi}\}$ with the corresponding quasienergies symmetrically distributed around $\delta$ and $\delta+\pi$ (due to SUB). Of course, for $\omega=\delta$, or $\omega=\delta+\pi$, it can happen that $\{ \ket{\Psi}\sim\Omega \ket{\Psi}\}$ and there is a single state instead of a dublet. It turns out that this is exactly the case of our edge states. Such single states cannot change their energy as long as the PHS is not broken.

\textit{Translation symmetry (TS)}.---Topological properties are defined for translation invariant walks ($\theta_x=\theta$). TS allows us to work in the quasimomentum basis $\ket{k} = \sum_x e^{ikx} \ket{x}$ with $k$ restricted to the first Brilloiun zone (BZ) $-\pi \leq k \leq \pi$. 
The effective Hamiltonian can be written as
\begin{equation}\nonumber
    H(\theta) = \oint_{BZ} \left(\ket{k}\bra{k}\otimes \bar{H}_k \right)dk,
\end{equation}
where $\bar{H}_k= \delta I+\omega_k {\bf {n}_k}\cdot\boldsymbol{\sigma}$ is the so-called Bloch Hamiltonian and
$\boldsymbol{\sigma} = (\sigma_x,\sigma_y,\sigma_z)$ is the vector of Pauli operators.
The normalized vector $ {{\bf{n}}_k}$ is given by
\begin{equation}\nonumber
    {{\bf{n}}_k} = \frac{\left\{\sin\theta\sin(k-\alpha') , -\sin\theta\cos (k-\alpha'), \cos\theta\sin (k-\alpha)\right\}}{\sin\omega_k},
\end{equation}
where $\alpha' = \alpha+\beta$ and $\omega_k$ (in the range $0\leq\omega_k\leq \pi$) is defined by equation
\begin{equation}\label{omega}
    \cos\omega_k = \cos\theta \cos (k-\alpha).
\end{equation}
The quasienergies are $\omega_{k\pm} =\delta \pm \omega_k$,
whereas the corresponding eigenstates are $\rho_{k\pm}=\frac{1}{2}(I \pm {{\bf {n}}_k}\cdot\boldsymbol{\sigma})$.
\begin{figure}[t]
\centering
\includegraphics[width=8cm]{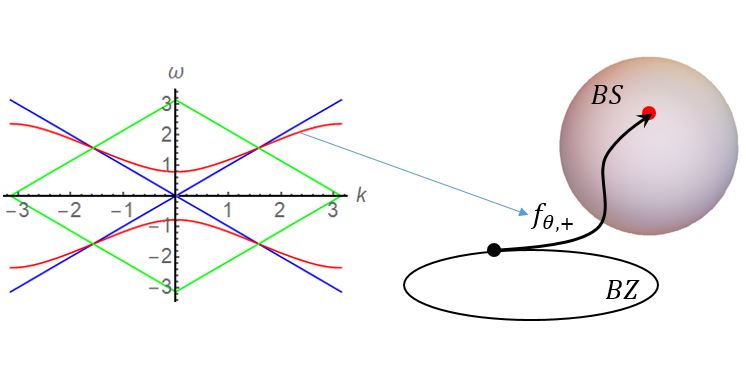}
\caption{The band structure of the basic DTQW (left). A typical gapped case for $\theta \in T$ (red line) and two gapless cases $\theta=0,\pi$ (blue and green lines). The graphical representation of the map $f_{\theta,+}$ corresponding to the upper band of the gapped spectrum (right).}
\label{fig1}
\end{figure}
It follows from the above equation (see Fig.\ref{fig1}) that the Hamiltonian is gapped (one gap around $\delta$ and one around $\delta+\pi$), with the exception of $\theta=0,\pi$, in which case both gaps close. Let $T$ be a subset of $\theta$ for which the Hamiltonian exhibits gaps. $T$ is considered to be a disconnected topological space with the following partition $T=T_{\theta>0}\cup T_{\theta<0}$, where $T_{\theta>0}=\{\theta |0<\theta<\pi\}$ and $T_{\theta<0}=\{\theta |,-\pi<\theta<0\}$.
This allows us to speculate that $T_{\theta>0}$ and $T_{\theta<0}$ may support distinct topological phases.

\begin{figure}[t]
\centering
\includegraphics[width=8cm]{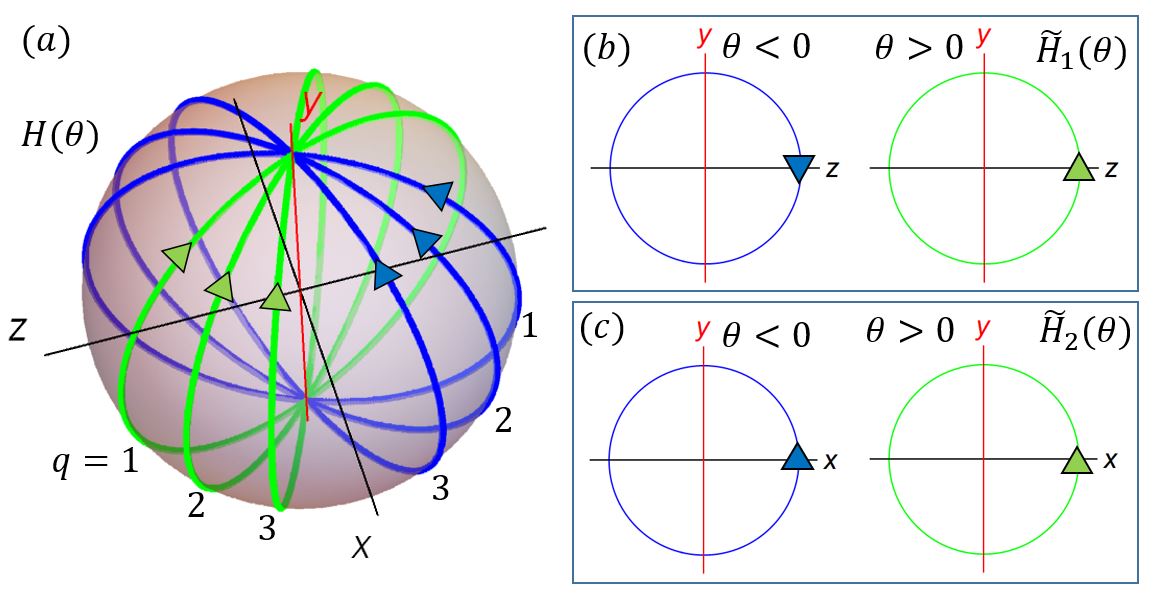}
\caption{Images of BZ given by $f_{\theta,+}(k)$ for a DTQW with $(\delta,\alpha,\beta)=(0,0,0)$. (a) Images for $H(\theta)$,  (b) images for $\Tilde{H}_1(\theta)$ and (c) images for $\Tilde{H}_2(\theta)$. Green and blue lines correspond to $0<\theta<\pi$ and $-\pi<\theta<0$, respectively. The parameter $q$ denotes subsequent values of $\theta=\pm q \frac{\pi}{8}$. The arrows indicate how the images of BZ wind when BZ is swept from $-\pi$ to $\pi$.}
\label{fig2}
\end{figure}

Our main mathematical tool will be a unique map from BZ to Bloch sphere (BS) $ f_{\theta,\pm}: BZ \to  BS$ defined as $f_{\theta,\pm}(k) = {\bf n}_{k \pm}$.
Our goal is to infer the properties of the DTQW directly from the topological properties of $f_{\theta,\pm}(k)$. In particular, we expect that the topological equivalence of these maps (given by the notion of homotopy \cite{Kitagawa:2010wv} should induce the partition of the parameter space into well defined (topological) phases.
An example of such mapping is presented in Fig.\ref{fig2} a. Note that ${\bf n}_k=-{\bf n}_{k+\pi}$. It follows that $f_{\theta,\pm}$ maps BZ to the great circles of BS. The arrows in Fig.\ref{fig2} indicate how the images of BZ wind when BZ is swept from $-\pi$ to $\pi$.

\textit{Parity symmetry (PS)}.---The translation invariant $H(\theta)$ exhibits an additional kind of unitary symmetry. Let us first define a generalized parity operator
\begin{equation}\nonumber
    \Pi_{\alpha} = \sum_x e^{-i2\alpha x}\ket{-x}\bra{x}=\sum_k \ket{2\alpha-k}\bra{k}.
\end{equation}
which can be perceived as a mirror inversion through ${k=\alpha}$.  In addition, let $P = \Pi_{\alpha}\otimes \bar{P}$
where $\bar{P} =i {\bf n}_{\beta} \boldsymbol{\sigma}$
with ${\bf n}_{\beta}=(sin \beta,cos \beta,0)$. $P$ is a unitary symmetry of the evolution operator $P U P^{-1} = U$.
We call it a PS of $H(\theta)$, since $P H(\theta) P^{-1} = H(\theta)$.
Note that $P$ does not depend on $\theta$ and is a symmetry of the whole family $S_{(\delta,\alpha,\beta)}$ of gapped Hamiltonians $H(\theta)$ defined for $\theta \in T$ and a fixed triple $(\delta,\alpha,\beta)$.

\textit{Chiral symmetry (CS)}.---For $\beta=0$ and $\theta \in T$ we define the vector (see Kitagawa et al. \cite{Kitagawa:2010vm}) ${\bf m}_\theta = \left\{\cos\theta,0,-\sin\theta\right\}$
orthogonal to ${{\bf{n}}}_k$ for any $k$. It follows that  $ \bar{\Gamma}(\theta) = e^{-i\frac{\pi}{2}{\bf m}_{\theta}\cdot \boldsymbol{\sigma}}$
leads to $\bar{\Gamma}(\theta) \bar{H}_k \bar{\Gamma}(\theta)^{-1}
    = -\bar{H}_k$,
so the effective Hamiltonian posses CS. However, $\bar{\Gamma}(\theta)$ cannot be regarded as CS of the whole family $H(\theta)$ due to $\theta$-dependence (see \cite{Asboth:2012ub}). Note, that an analysis of topological properties focuses on a family of Hamiltonians that dependent on some continuous parameter. It was shown in \cite{Asboth:2012ub} that there is no CS imposed by a single ($\theta$-independent) unitary operator $\Gamma=I \otimes \bar{\Gamma}$ which fulfills $\bar{\Gamma} \bar{H}_k(\theta) \bar{\Gamma}^{-1} = -\bar{H}_k(\theta)$
for all $\theta \in T$.

\begin{figure}[t]
\centering
\includegraphics[width=8cm]{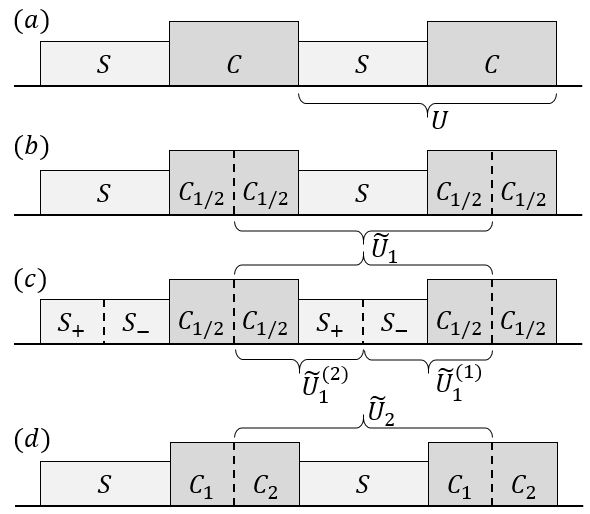}
\caption{A sequence of unitary operations that make up a basic DTQW: (a) an original sequence, (b, c, d) different decomposition of the original sequence corresponding to the time-shifted DTQW.}
\label{fig3}
\end{figure}

\textit{Unitary equivalence}.---Although $\bar{\Gamma}(\theta)$ depends on $\theta$, its eigenvalues $\pm i$ do not. Therefore, there must exist a unitary $\bar{V}(\theta)$ such that $\bar{\Gamma}(\theta) =  \bar{V}(\theta)^{-1} \bar{\Gamma}\bar{V}(\theta)$
where $\bar{\Gamma}=i \ \boldsymbol{\gamma}\cdot \boldsymbol{\sigma}$ does not depend on $\theta$ (${\boldsymbol \gamma}$ is some fixed normalized vector).
We define a family $\Tilde{S}_{(\delta,\alpha,\beta)}$ of modified Hamiltonians (unitarily equivalent to the original ones)
\begin{equation}\nonumber
    \Tilde{H}(\theta) = V(\theta)H(\theta) V(\theta)^{-1} 
\end{equation}
where $V({\theta})=I \otimes \bar{V}({\theta})$. The family of modified Hamiltonians has a desired $\theta$-independent CS $\Gamma \Tilde{H}(\theta) \Gamma^{-1} = -\Tilde{H}(\theta)$
where $\Gamma=I \otimes \bar{\Gamma}$.
The unwanted $\theta$-dependence of CS has been moved to the unitary transformation $V(\theta)$. For example, if $\bar{V}_1(\theta) = e^{i \frac{\theta}{2} \sigma_y}$
then the family $\Tilde{H}(\theta)$ has $\theta$-independent CS  $\Gamma_1=I \otimes \bar{\Gamma}_1$, where $\bar{\Gamma}_1 = \sigma_x$ (with omitted irrelevant imaginary unit).
The transformation $\bar{V}(\theta)$ rotates the BZ-to-BS maps $f_{\theta,\pm}(k)$ such that they all lie in the YZ-plane (hence the CS). After the transformation the maps for $\theta > 0 $ and $\theta < 0$ (see Fig.\ref{fig2} b) wind in the opposite directions. We can formalize this observation using the concept of the winding number (around X-axis) that takes the value $-1$ for $\theta>0$ and $+1$ for $\theta <0$. 


\textit{Possible issues}.---It is tempting to characterize the topological phases of $H(\theta)$  with the help of winding numbers obtained for $\Tilde{H}(\theta)$. Unitary equivalence guarantees that both Hamiltonians posses the same spectrum. However, it is known that topological properties cannot be deduced solely from the spectrum. Therefore, it is not trivial to ask if it is possible to make statements about topological properties based on unitary equivalence (\cite{Tarasinski:2014tv}).

To explore some potential problems let us consider another unitary operator $\bar{V}_2(\theta) = e^{\frac{i}{2} (\theta-sgn(\theta)\frac{\pi}{2}) \sigma_y}$.
The operator $\Tilde{H}_2(\theta) = V_2(\theta)H(\theta) V_2(\theta)^{-1} $ has CS imposed by $\bar{\Gamma}_2 = \sigma_z$
The corresponding BZ-to-BS maps are presented in Fig.\ref{fig2} c. All of them lie in the YX-plane, but this time the winding numbers (around Z-axis) are all the same and do not depend on $\theta$. Therefore, we have two families of Hamiltonians $\Tilde{S}_{1,(\delta,\alpha,\beta)}$ and $\Tilde{S}_{2,(\delta,\alpha,\beta)}$, both unitarily equivalent to the original family (and to each other), but exhibiting different topological properties.

\textit{Time-shifted frames}.---Asb\'oth and Obuse introduced an interesting method to construct CS DTQWs \cite{Asboth:2013aa}. Its advantage stems from an intuitive physical interpretation of the unitary equivalence. Let us consider $\bar{C} = e^{i \theta \sigma_y}$ 
and define $C_{1/2}=I \otimes \bar{C}_{1/2}$,
where $\bar{C}_{1/2}= e^{i \frac{\theta}{2} \sigma_y}$.
The following unitary evolution operator
\begin{equation}\nonumber
    \Tilde{U}_1=V_1(\theta) U V_1(\theta)^{-1}=C_{1/2} \ S \ C_{1/2},
\end{equation}
looks like $U$ shifted in time (see Fig.\ref{fig3} a, b). 
Obviously $\Gamma_1$ is CS of $\Tilde{U}_1$.

A natural interpretation in terms of a time-shifted frame seems to favour $V_1(\theta)$ over $V_2(\theta)$. This observation is, however, based on some hidden assumption. If the coin operator is generated by some constant Hamiltonian $H_C$ (during some period $T$) $\bar{C} = e^{i \theta \sigma_y}=e^{-i H_c T}$
then it is natural to consider a half of the period $\frac{T}{2}$ and decompose $C$ as 
\begin{equation}\label{dec}
    \bar{C} = e^{-i H_c T/2}e^{-i H_c T/2}=\bar{C}_{1/2}\bar{C}_{1/2}.
\end{equation}
As a consequence $ \bar{V}_1(\theta) = e^{i \frac{\theta}{2} \sigma_y}=e^{-i H_c T/2}$
can be understood as a time shift. However, DTQW is defined by the coin operator $\bar{C}$ without any further information about its implementation. Therefore, the decomposition (\ref{dec}) is not unique. One can as well consider a decomposition (see Fig. \ref{fig3}d) $\bar{C} = \bar{C}^{(1)}\bar{C}^{(2)}$,
where $\bar{C}^{(1)}\neq\bar{C}^{(2)}$. For example, let $\bar{C}^{(1)}=\bar{C}_{1/2} \bar{C}_s$ and $ \bar{C}^{(2)}=\bar{C}_s^{-1}\bar{C}_{1/2} $
where $ \bar{C}_s=e^{i \frac{\pi}{4} sgn(\theta) \sigma_y}$.
This leads to the following time-shifted DTQW $\Tilde{U}_2=C^{(2)}  S C^{(1)}$.
We obtained two DTQWs, both corresponding to the original DTQW, with different topological properties.

\begin{figure}[t]
\centering
\includegraphics[width=8cm]{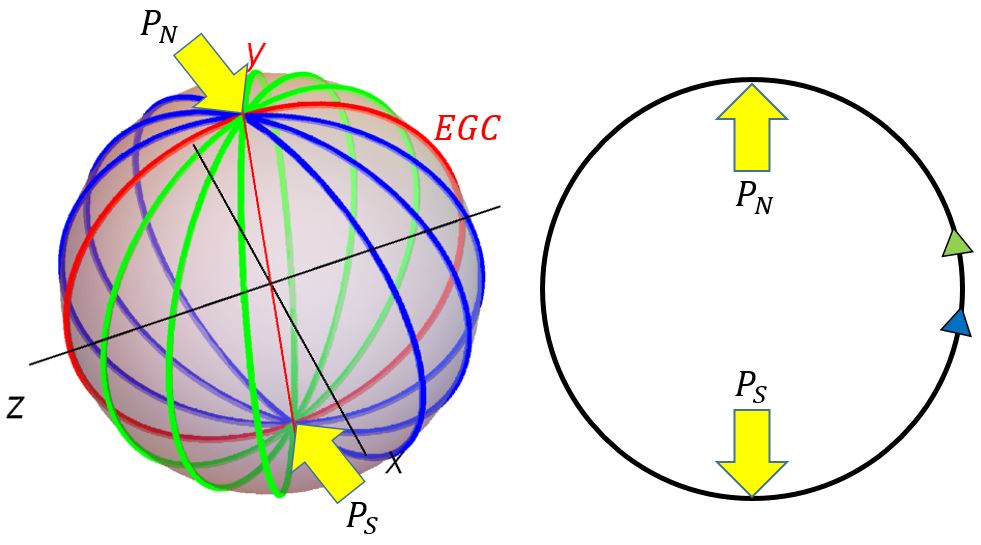}
\caption{(left) Bloch sphere with depicted poles and an excluded great circle (red). (right) A circle with depicted poles, that is homotopic equivalent to the manifold $M_T$. Green and blue arrows indicate how the images of BZ wind for positive and negative $\theta$, respectively.}
\label{fig4}
\end{figure}

\textit{Homotopic maps}.---Let us look closer at the manifold $M_T$ defined as a union of BZ images generated by the map $f_{\theta,+}(k)$ 
\begin{equation}\nonumber
    M_T=\bigcup_{\theta \in T} \ \{{\bf n}_{k+}=f_{\theta,+}(k) |k \in BZ\}.
\end{equation}
Note that $f_{\theta,-}(k)$ does not need to be considered separately since it forms the same manifold. It turns out that $M_T$ is a proper subset of BS, hence instead of $ f_{\theta,\pm}: BZ \to  BS$ we focus on topological properties of 
\begin{equation}\nonumber
    f_{\theta,\pm}: BZ \to  M_T.
\end{equation}

$M_T$ consists of two hemispheres connected by just two points (the poles). The union of these hemispheres is equivalent to the whole BS without a great circle (see Fig.\ref{fig4}). The connecting poles are located at the intersection of the XY-plane with the excluded great circle (EGC). Therefore, $M_T=(BS\backslash EGS) \cup P_N\cup P_S$.
For the DTQW with $\bar{C} = e^{i \theta \sigma_y}$ the EGC lies in the $YZ$-plane and the coordinates of the north and the south poles are given by $P_N=(0,1,0)$ and $P_S=(0,-1,0)$, respectively.

Now we make a crucial observation that the manifold $M_T$ is a homotopy equivalent to a circle (its fundamental group is isomorphic to $\mathbb{Z}$, see Fig.\ref{fig4}). 
The consequence of the above is that the winding numbers of the maps $f_{\theta,\pm}$ are well defined, even in the absence of CS. The crucial property of these winding numbers is that they are the same for all $\theta$ (see Fig.\ref{fig4}), therefore all the corresponding maps are homotopic. Moreover, any transformation that changes the winding number (like $V_1(\theta)$) is not compatible with the topological structure of the model. Hence, we have to find an alternative way of identifying different topological phases.

\begin{figure}[t]
\centering
\includegraphics[width=8cm]{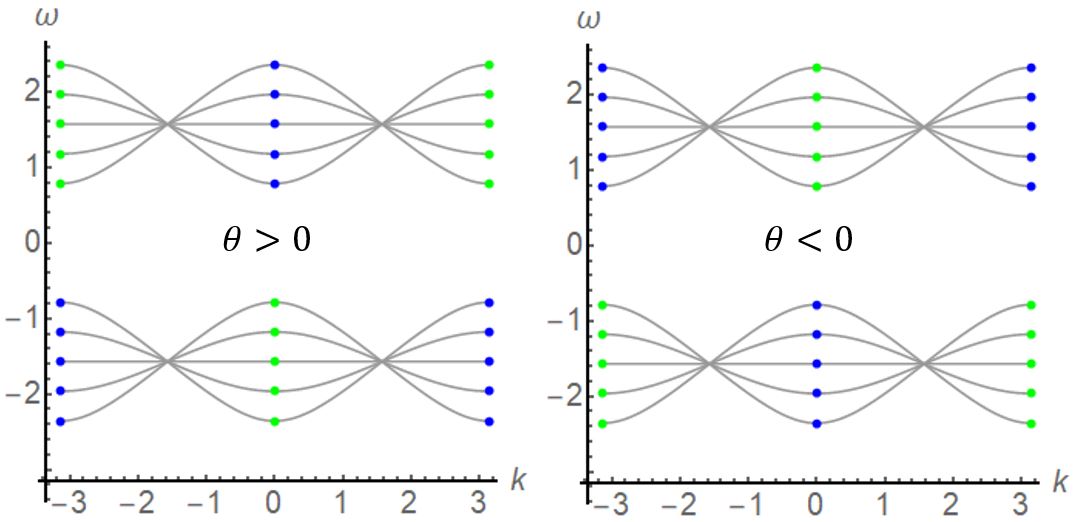}
\caption{Examples of band structures for $|\theta|=q \frac{\pi}{8}$ ($q=2,3,4,5,6$). Green dots correspond to ${\bf n}_\beta$ and blue ones to $-{\bf n}_\beta$.}
\label{fig5}
\end{figure}

Let us return for a moment to parity and translation symmetries. PS does not commute with TS. However, there are two special points in BZ $k_j=\alpha+j \pi$ ($j=0,1$) for which $H_{k_j}$ is $\bar{P}$ invariant $\bar{P} H_{k_j} \bar{P}^{-1} = H_{k_j}$.
Significance of these points (in the case $\alpha=0$) was recognized in \cite{Asboth:2012ub}, where it was highlighted that the qusienergy gaps can close at $k_j$. Here we focus on yet another important property of these points (see Fig.\ref{fig5}). The commutation relation $[\bar{P}, H_{k_j}]=0$ guarantees that the eigenvectors of $H_{k_j}$ must be also the eigenvectors of $\bar{P}$. Therefore, they do not depend on $\theta$. These eigenvectors are denoted $\pm {\bf n}_{\beta}$. Their existence implies a specific form of the manifold $M_T$. Namely, the vectors $\pm {\bf n}_{\beta}$ are positions of the poles, whereas EGS lies in the plane that includes both, the Z-axis and the poles. 

The position of the poles (see Fig.\ref{fig5}) allows to pinpoint a difference between $\theta>0$ and $\theta<0$. We have
\begin{equation}\nonumber
    P_N =f_{\theta>0,+}(k_1)=f_{\theta<0,+}(k_0),
\end{equation}
and
\begin{equation}\nonumber
    P_S =f_{\theta>0,+}(k_0)=f_{\theta<0,+}(k_1).
\end{equation}
It follows that it is impossible to continuously transform $f_{\theta>0,+}(k)$ into $f_{\theta<0,+}(k)$ without a violation of PS.

\begin{figure}[t]
\centering
\includegraphics[width=9cm]{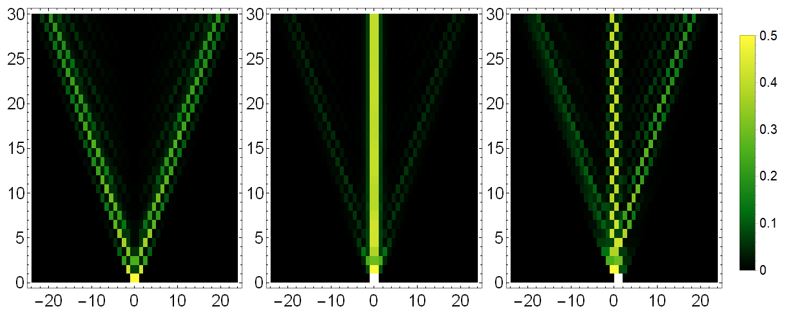}
\caption{An example DTQW dynamics for $(\delta,\alpha, \beta)=(0,0,\frac{\pi}{2})$ at the interface of two distinct DTQWs corresponding to $\theta_{1,2}=\mp \frac{\pi}{4}$. The initial state is centered at the interface and: (left) is orthogonal to both edge states, (middle) has nonzero overlap with one edge state and is orthogonal to the other one, (right) has nonzero overlap with both edge states.}
\label{fig6}
\end{figure}

\textit{Relative homotopy and corresponding invariant}.---To formalize the above observation we propose to use the concept of relative homotopy. The functions $f$ and $g$ are homotopic relative to a point $a^*$ if and only if they are homotopic and $f(a^*)=g(a^*)$.
For the clarity of presentation let us recall the definition of homotopy. Let $A$ and $B$ be topological spaces and let $f$ and $g$ be continuous functions from $A$ to $B$. Moreover, let $I$ be a unit interval $[0,1]$. We say that $f$ and $g$ are homotopic if and only if there exists a continuous function $ h: A \times I  \to B$
such that for all $a \in A$ $h(a,0)=f(a)$
and $h(a,1)=g(a)$.
We can now examine whether the two maps $f_{\theta 1,+}(k)$ and $f_{\theta 2,+}(k)$ are homotopic relative to $k_0$ and $k_1$. In each case we obtain a yes/no answer. The idea is to use these answers as topological invariants. We can expect $\mathbb{Z}_2\otimes\mathbb{Z}_2$ classification of topological phases. It turns out that maps with the same sign of $\theta$ are homotopic relative to both $k_0$ and $k_1$, whereas maps with the opposite sign are not (see Fig.\ref{fig4}). Each of two points corresponds (for $\theta=0$ or $\theta=\pi$) to closing of one of the gaps at quasienergy $\delta$, or $\delta+\pi$. We thus expect two edge states (one in each gap) at the interface between distinct topological phases. Let us remind that such single states are perfectly protected by PHS. 

\textit{Sharp interface}.---Let us show the exact form of two edge states at the interface of two distinct DTQWs. Without a loos of generality we assume that $\theta_x=\theta_1<0$ for $x<0$ and $\theta_x=\theta_2>0$ for $x\geq0$. The corresponding edge states are (for $\eta=0,\pi$)
\begin{equation}\nonumber
    \ket{\Psi_\eta} =\frac{1}{\sqrt{N}} \sum_x e^{i \eta x} \ket{x}\otimes(a_x\ket{\rightarrow}+b_x\ket{\leftarrow})
\end{equation}
where
$a_x=A_j^x$ and $b_x=-e^{-i(\alpha+\beta)}A_j^{x+1}$ with $j=1,2$ for $x<0$ and $x\geq0$ respectively. $A_j=e^{i \alpha} \frac{1-\sin \theta_j}{\cos \theta_j}$ whereas $N=\sin^{-1} \theta_2-\sin^{-1} \theta_1$. In Fig. \ref{fig6} we present an example DTQW dynamics for $(\delta,\alpha, \beta)=(0,0,\frac{\pi}{2})$ and $\theta_{1,2}=\mp \frac{\pi}{4}$. We consider three different initial states that are centered at the interface. In the first case the initial state is orthogonal to both edge states and we observe a ballistic departure of the wave packet from the interface. In the second case the initial state has nonzero overlap with one of the edge states, therefore we observe a localization at the interface. Finally, in the last case the initial state has nonzero overlap with both edge states, hence we observe a localization at the interface and the emergence of the interference pattern between the two edge state constituents. 

\textit{Summary and outlook}.---We reconsidered topological properties of basic DTQWs and proved that it is possible to infer them directly from the mapping of the Brillouin zone to the Bloch Hamiltonian. We showed that it is convenient, if not necessary, to use the concept of relative homotopy and introduced a topological invariant based on it. This allowed us to identify two topological phases of homogeneous DTQWs. Moreover, we were able to identify two localized edge states (one in each quasienergy gap) at the interface between two distinct phases of the inhomogenous DTQW for any (fixed) values of $(\delta,\alpha, \beta)$. We also found PHS that protects these states. Finally, we provided an exact form of edge states in the case of a sharp boundary between distinct topological phases.
\\
\\\indent
This research is supported by the Polish National Science Centre (NCN) under the Maestro Grant no. DEC-2019/34/A/ST2/00081.  J.W. acknowledges support from IDUB BestStudentGRANT (NO. 010/39/UAM/0010). M.K. is supported by Foundation for Polish Science (IRAP project, ICTQT, con-
tract no.2018/MAB/5, co-financed by EU within Smart Growth Operational Programme).

\end{document}